\documentclass[aps,prb,twocolumn,superscriptaddress,floatfix,showpacs,10pt,letterpaper]{revtex4-1}

\allowdisplaybreaks[1]
\pdfoutput=1

\usepackage{amsthm}
\usepackage{euscript}
\usepackage{epstopdf} 

\begin{document}

\newcommand{\ket}[1]{|#1\rangle}
\newcommand{\bra}[1]{\langle#1|}

\newcommand{\braket}[2]{\langle #1 | #2 \rangle}
\newcommand{\ketbra}[2]{|#1 \rangle \langle #2 |}

\begin{titlepage}

\title{
Topological degeneracy (Majorana zero-mode) and 1+1D fermionic topological order\\
in a magnetic chain on superconductor via spontaneous $Z_2^f$
symmetry breaking
}

\author{Joel Klassen} 
\affiliation{Department of Physics, University of Guelph, 50 Stone Road East, Guelph, Ontario, Canada}
\affiliation{Institute for Quantum Computing, 200 University Avenue West, Waterloo, Ontario}
\author{Xiao-Gang Wen} 
\affiliation{Department of Physics, Massachusetts Institute of Technology, Cambridge, Massachusetts 02139, USA}
\affiliation{Perimeter Institute for Theoretical Physics, Waterloo, Ontario, N2L 2Y5 Canada}

\begin{abstract} 


We study a chain of ferromagnetic sites, ie nano-particles, molecules or atoms, on a substrate of fully gapped superconductors.  We find that under quite realistic conditions, the fermion-number-parity symmetry $Z_2^f$ can spontaneously break.  In other words, such a chain can realize a 1+1D fermionic \emph{topologically ordered} state and the corresponding two-fold \emph{topological degeneracy} on an open chain.  Such a topological degeneracy becomes the so called \emph{Majorana zero mode} in the non-interacting limit.

\end{abstract}


\maketitle

\end{titlepage}


\section{Introduction}
\label{intro}

Recently, there has been a strong experimental and theoretical effort to search for the
\emph{Majorana zero
mode}\cite{MZF1236,NDL1446,W0914,F1024,S1193,B1313,JR8181,FK0807,L0981,LSD1001,CEA1119,KSY1305,DRT1412,PPG1451}
(which is often wrongly and misleadingly referred to as the
\emph{Majorana fermion}). However, the Majorana zero mode is actually a feature of systems
of non-interacting fermions.  So, strictly speaking, the Majorana zero mode does
not exist in any realistic systems where electrons interact.  In fact,
what people are truly interested in is not the Majorana zero mode, but
\emph{topological degeneracy}.  Topological degeneracy is the ground state
degeneracy of a gapped Hamiltonian system in large system size limit, which is
robust against any perturbations that can break any
symmetry.\cite{W8987,WN9077} Topological degeneracy is a sign of
\emph{topological order}.\cite{W9039,KW9327} So the search for ``Majorana
fermions''\footnote{In fact, Majorana fermions had already been found 50 years
ago in superconductors, but under a different name, the Bogoliubov
quasiparticles.} is actually the search for topological degeneracy in
topological order.

Topological order is a new kind of order in gapped quantum systems that extends
beyond the Landau symmetry breaking description.\cite{W8987,WN9077,W9039,KW9327}
For bosonic systems, topological order can only exist in 2+1-dimensions and
higher.\cite{VCL0501,CGW1038,CGW1107}  Bosonic topological order can lead
to topological degeneracy if the system lives on a torus\cite{W8987,WN9077} or
has several disconnected boundaries.\cite{WW1263,KK1251}  For fermionic
systems, fermionic topological order\cite{GWW1017} can even exist in
1+1D.\cite{K0131}  Such 1+1D fermionic topological order can lead to a two-fold
topological degeneracy if the system lives on an open line segment.\cite{K0131}

Since fermionic systems always have a fermion-number-parity (FNP) symmetry
$Z_2^f$ which can never be explicitly broken, the above 1+1D fermionic
topological order can be viewed as a spontaneous symmetry breaking order of the
FNP symmetry $Z_2^f$ (at least when the systems live on an open line
segment).\cite{Sac11,CGW1128,GW1248}  The above mentioned  two-fold topological
degeneracy is nothing but the two-fold degeneracy of the $Z_2^f$ symmetry
breaking.  As a result, we can study the 1+1D fermionic topological order and
its topological degeneracy on an open line using Landau symmetry breaking
theory.

In this paper, we will consider a chain of ferromagnetic nano-particles or
ferromagnetic molecule/atoms on a substrate of superconductor.  We find that
under quite realistic conditions, the FNP symmetry breaking state can appear
(or 1+1D fermionic topologically ordered state can appear), which will lead to
an experimental realization of topological degeneracy.  Our approach also
allows us to understand the relevant energy scales: the energy splitting $\del
E_{eo}$ between the states of even and odd electrons on a nano-particle, the
hopping amplitude $t_{ij}$ between nano-particles, and the Josephson coupling
$J_i$ between the superconducting substrate and the nano-particle.  We also
understand when the topological degeneracy can be observed at higher
temperatures: (1) $|t_{ij}| \sim |J_i|$  are large, (2) $|t_{ij}| \gtrsim \del
E_{eo}$, and (3) the phase of $J_it_{ij}^2J_j^*$ is not zero.  

Chains of magnetic nano-particles on a substrate of fully gapped superconductor
have been studied theoretically by mapping the system to an effective free
Majorana fermion chain.\cite{CEA1119,KSY1305} In this paper, we study a
different parameter regime which leads to a different effective theory.  Chains
of magnetic iron (Fe) atoms on a substrate of superconducting lead (Pb) were
recently studied experimentally in \Ref{NDL1446}, where features of the Majorna
zero mode was found.\cite{DRT1412,PPG1451}

\section{The model}

We will use the following effective Hamiltonian to describe a chain of
magnetic dots on a substrate of fully gapped superconductor 
\begin{align}
\label{Hf}
H &=\sum_i [ 
t \hat c_{i+1}^\dag \hat c_i
+J \hat c_i \hat c_i
+ h.c.]
\nonumber\\
&
+ \sum_i \Big[U (\hat n_i- n_0)^2+\Del \frac{(-)^{\hat n_i}-1}{2}\Big]
,
\end{align}
where $\hat n_i$ is the fermion number operator and $\hat c_i$ is the effective
(spinless) fermion operator acting on the Hilbert space $\cV_i$ on site-$i$.
$\cV_i$ is formed by states of $n$-fermions, $n=0,\pm1,\pm2,$ etc and $\hat
n_i$ and $\hat c_i$ satisfy
\begin{align}
&
\{\hat c_i,\hat c_j\}= \{\hat c_i,\hat c_j^\dag\}=[\hat c_i,\hat n_j]=0,\ \ i\neq j,
\nonumber\\
&\
\hat c_i|n\>=|n-1\>.
\hat n_i|n\>=n|n\>.
\end{align}
Note that 
the eigenvalue of $\hat n_i$ can be any integer $n$,  and $\hat c_i$ is not the
standard fermionic operator.

\begin{figure}[b]
\includegraphics[width=1.5in]{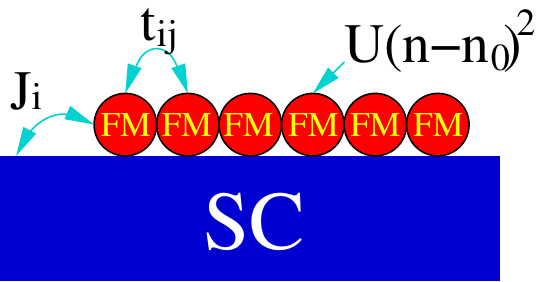}
\caption{\label{device} The geometry of the device: a chain of ferromagnetic dots on a fully gapped superconductor. }
\end{figure}

In our effective Hamiltonian \eq{Hf} (see Fig. \ref{device}), $\Delta$ is the induced pairing energy on
the magnetic dot. $U$ is the effective Coulomb repulsion on the dot.  The
effect of chemical potential or gate voltage is summarized by $n_0$.  $t$ is
the electron hoping amplitude between neighboring dots and $J$ is the Josephson
coupling between the dots and the superconducting substrate.  Since the dots
are magnetic, the spin degree of freedom is assumed to be frozen.  We also
assume the dots are ferro- or anti-ferro-magnetically ordered, so that there is
no spatial dependence in $t$.

To understand the phase diagram of the above interaction 1+1D fermionic system on
an \emph{open chain}, we may perform a Jordan-Wigner transformation
\begin{align}
\hat c_i^{\dagger}=&\hat n_i^{+}\prod_{j<i}(-1)^{\hat{n}_j}&
\hat c_i=\hat n_i^-\prod_{j<i}(-1)^{\hat{n}_j},
\end{align}
where the action of these operators are as follows
\begin{align}
&\hat{n}_i\ket{n}=n\ket{n} \nonumber \\
&\hat n_i^+\ket{n}=\ket{n+1}\\
&\hat n_i^-\ket{n}=\ket{n-1} \nonumber 
\end{align}
Our bosonic effective Hamiltonian then takes the form
\begin{equation}\label{hamiltonianN}
\begin{aligned}
H&=\sum_{i} \Big[ U(\hat{n}_i-n_0)^2+\Delta \frac{(-1)^{\hat{n}_i}-1}{2}
\\
&+ (J \hat n_i^{+}\hat n_i^{+}+h.c.)+(t \hat n_i^{+}(-1)^{\hat{n}_i}\hat n_{i+1}^-+h.c.) \Big]
\end{aligned}
\end{equation}
The FNP $Z_2^f$ transformation is generated by $(-)^{\sum_i
\hat n_i}$, which is a symmetry of the above effective Hamiltonian.

\section{The phase diagram} 
\label{phase}

\subsection{Small $t$ limit}

 \begin{figure}[tb]
 \includegraphics[width=1.in]{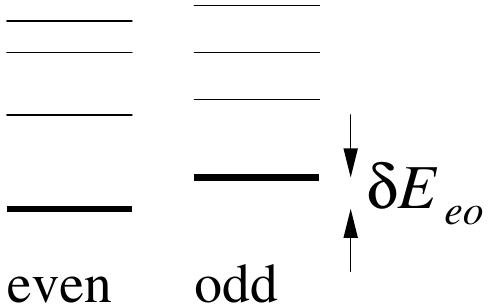}
 \caption{\label{dEeo} The many-body energy levels on a single dot, 
with even and odd electron numbers.}
 \end{figure}

When $t$ is small, we can solve the one-site Hamiltonian
\begin{align}
H_i=U(\hat{n}_i-n_0)^2+\Delta \frac{(-1)^{\hat{n}_i}-1}{2}
+(J \hat n_i^{+}\hat n_i^{+}+h.c.)  
\nonumber 
\end{align}
first.  Let us assume the the two lowest energy eigenstates of $H_i$ are formed by
one even-fermion state $|\up\>$ and one odd-fermion state $|\down\>$ (see Fig. \ref{dEeo}).
In this lowest energy subspace, $H_i$ becomes
$ H_i=h_z \si^z_i$,
where $\si^{x,y,z}_i$ are the Pauli matrices acting on $|\up\>,|\down\>$.
In the subspace $|\up\>,|\down\>$, $(-)^{\hat n_i}=\si^z_i$ and
$\hat n_i^+$ has a form
$ \hat n_i^+=\ee^{\ii\phi}(h_x\si^x_i+\ii h_y\si^y_i)$,
where $h_{x,y}\sim O(1)$ are real and positive.
Therefore, $H$ in \eqn{hamiltonianN} becomes
\begin{align}
\label{Hsi}
H&=\sum_{i} \Big[ h_z\si^z_i
+2\Re(t)h_xh_y ( \si^x_i\si^x_{i+1} +\si^y_i\si^y_{i+1})
\nonumber\\
&\ \ \ \ \  \ \ \ \ \ \ \ \ \ \ \
+2\Im(t) ( h_x^2\si^y_i\si^x_{i+1} - h_y^2\si^x_i\si^y_{i+1})
 \Big]
\end{align}

We can use mean-field theory to find the phase diagram of the above
spin-1/2 Hamiltonian by assuming a uniform spin order if $\Re(t)<0$:
\begin{align}
\v \si_i =
\cos(\phi ) \sin(\th) \v x
+\sin(\phi ) \sin(\th) \v y
-\cos(\th) \v z
\end{align}
The corresponding average ground state energy per site is given by  
\begin{align}
\frac{\langle H \rangle}{N}
&= -h_z\cos(\th) +
2\Re(t)h_xh_y \sin^2(\th)
\nonumber\\
&
+\Im(t)(h_x^2-h_y^2) \sin(2\phi) \sin^2(\th)
\end{align}
Assuming some typical values $h_x=2/3$, $h_y=1/3$, and $\Re(t)=\Im(t)$,
we have
\begin{align}
\frac{\langle H \rangle}{N}
&= -h_z\cos(\th) +
\frac{7}{9}\Re(t) \sin^2(\th), \ \ \ \ \sin(2\phi)=1,
\nonumber 
\end{align}
and the $Z_2^f$ symmetry breaking happens when
$ -\frac{\Re(t)}{h_z} > \frac{9}{14}$.

We note that when $t$ is real, the effective theory has a $U(1)$ symmetry
generated by $\ee^{\ii \th \sum_i \si^z_i}$, where $Z_2^f$ is part of the
$U(1)$. In this case $U(1)$ and $Z_2^f$ symmetry breaking cannot happen when we
include the quantum fluctuations beyond the mean-field theory.
Even when $t$ is complex, we still require $h_x-h_y$ to be large which
requires $|J|\gtrsim |h_z|$.
We conclude that $Z_2^f$ symmetry breaking or 1+1D topological order
can appear if\\
(1) the electron hopping $t_{ij}$ between dots is larger
than the energy splitting $\del E_{eo}=2h_z$ between states of even and odd electrons
on a dot, \\
(2) the Josephson coupling $J_i$ 
between the superconducting substrate and the dot satisfy $|J_i|\gtrsim \del E_{eo}$,\\
(3) the electron hopping amplitude $t_{ij}$ is complex, or more
precisely, the phase of the gauge invariant combination $J_i t_{ij}^2 J_j^*$ is not zero.

Note that we can tune the energy splitting between the states of even and odd
electrons to zero by tuning the gate voltage.  In this case, we only require
the electron hopping $t_{ij}$ to be larger than the fluctuation of the energy
splitting between even and odd states (caused by randomness).  In other words,
the electron hopping $t_{ij}$ should overcome the localization effect (at the
Josephson coupling energy scale).

\subsection{Mean-field theory for generic case}

In the small $t$ limit, only two states per dot are involved.  For large $t$,
we need to use the more general model \eq{hamiltonianN}, where many states on
each dot are included.  We can also employ a mean-field approximation for the
general model \eq{hamiltonianN} by assuming that the trial ground state of this
Hamiltonian takes the form
\begin{align}
&\ket{\psi}=\prod_j \ket{\psi_j}
\nonumber \\
&\ket{\psi_i}=\sum_i r_n^i \ee^{\ii \theta_n^i }\ket{n} , \ \ \ \ 
\sum_n  (r_n^i)^2=1 , \ \ \  r_n^i\geq 0.
\end{align}
Since the total phase of the quantum wave function is unphysical,
$\ket{\psi_i}$ is actually labeled by $(r_n^i,\Del \th_n^i)$ [not by
$(r_n^i,\th_n^i)$], where 
\begin{equation} 
\Delta \theta_n^i=\theta_{n+1}^i-\theta_n^i 
\end{equation}

\begin{figure}[t]
\includegraphics[width=3.9in]{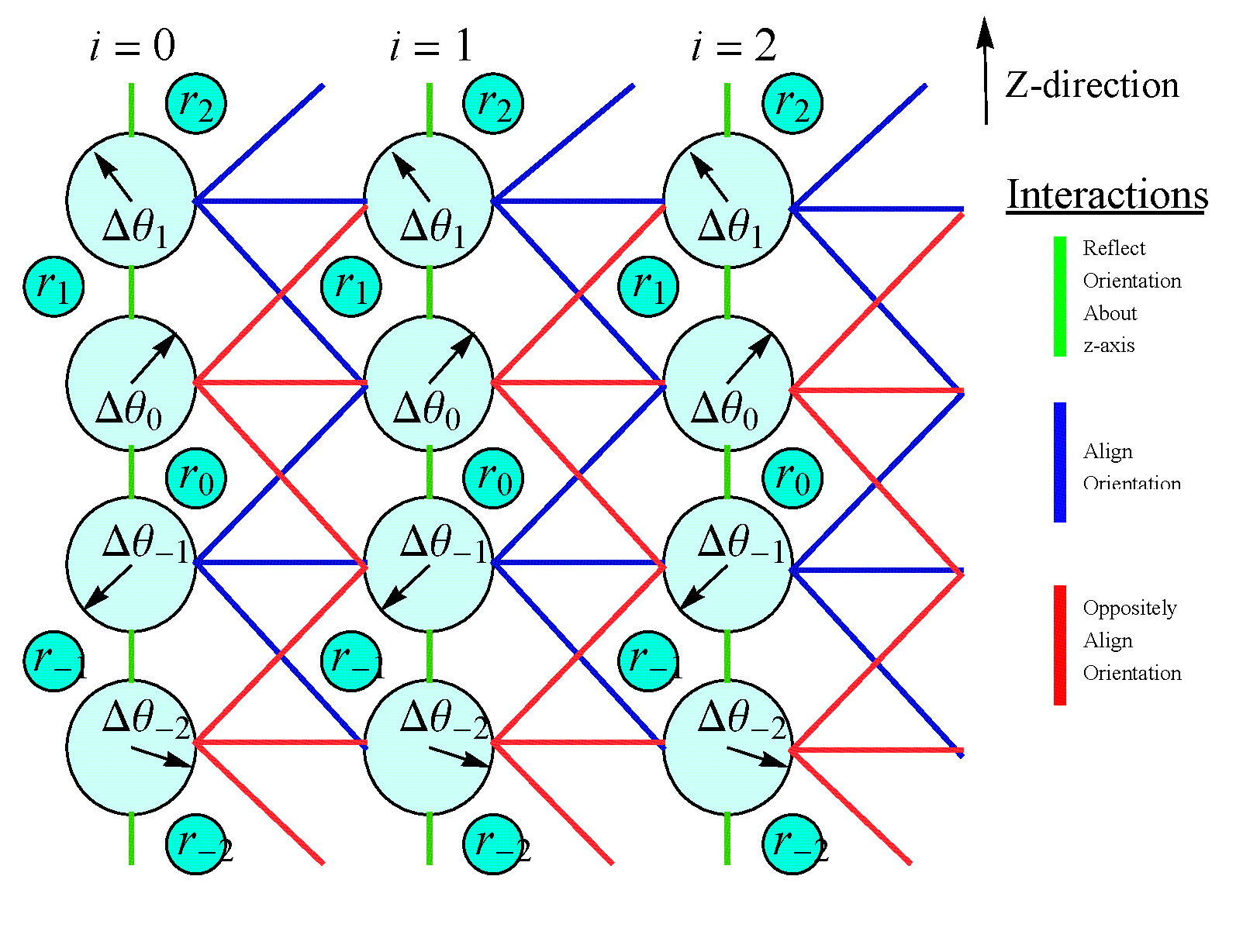}
\caption{\label{fig:model} Ground state energy minimization model. }
\end{figure}
\begin{figure}[t]
\includegraphics[width=2.4in]{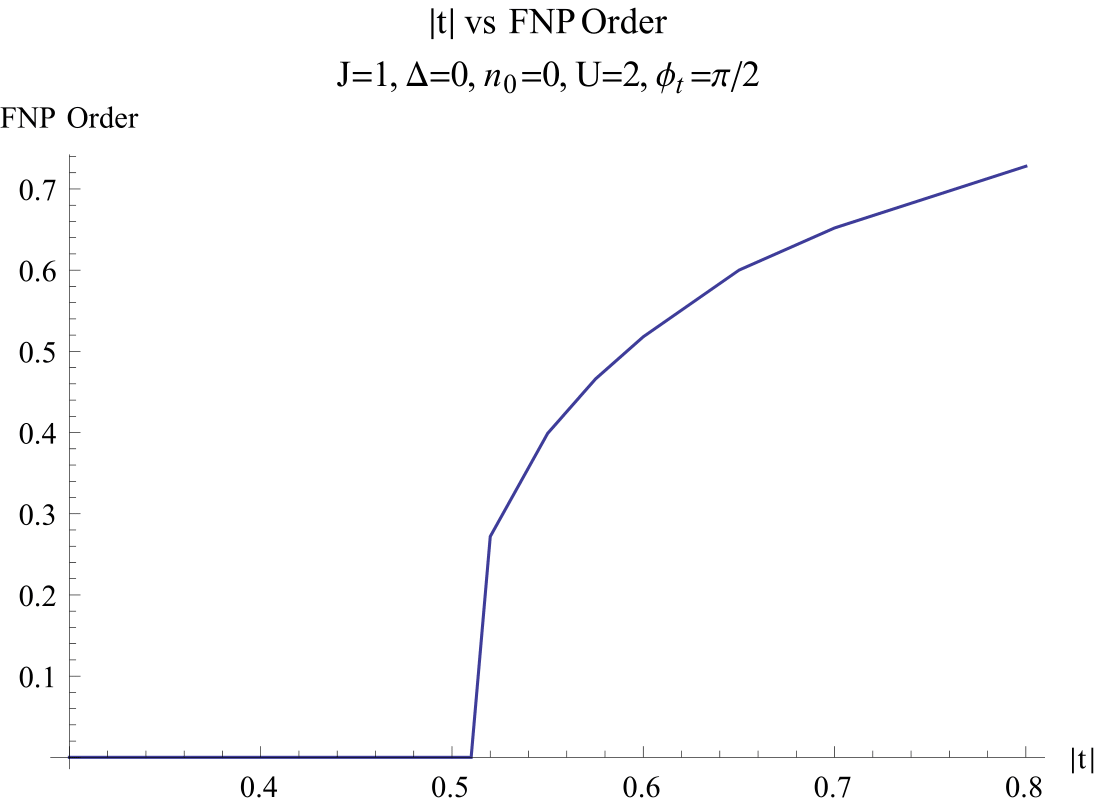}\\
\includegraphics[width=2.4in]{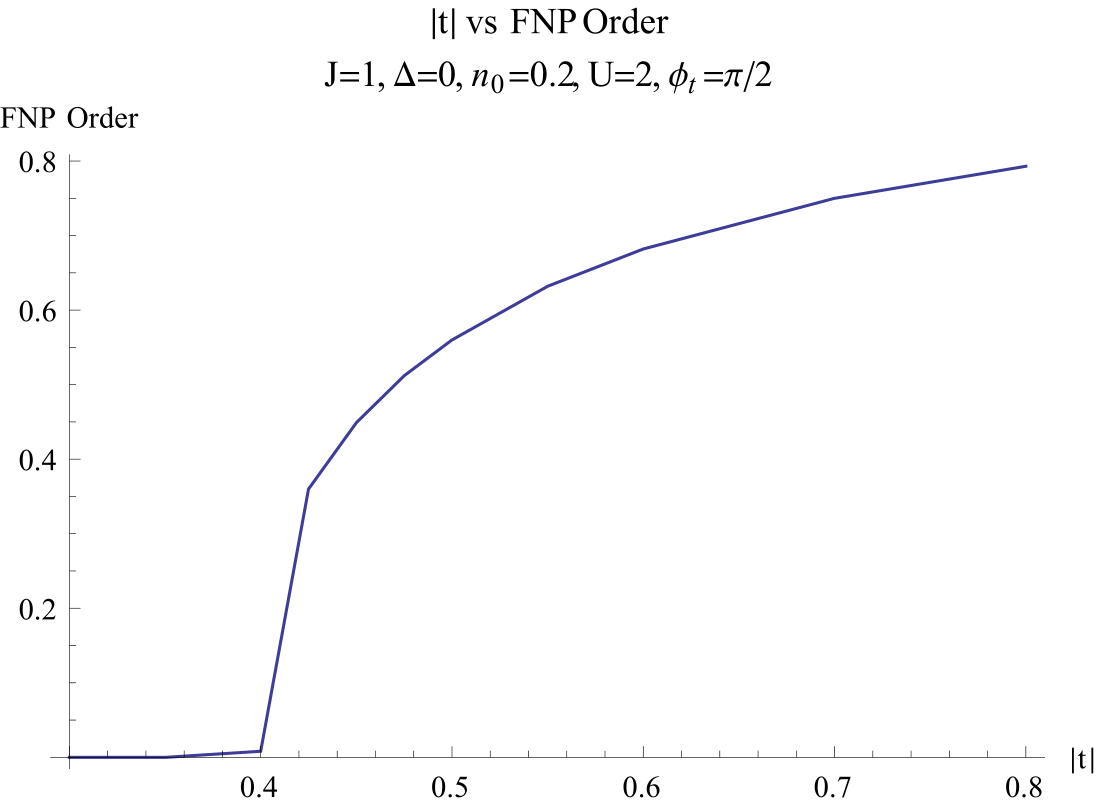}\\
\includegraphics[width=2.4in]{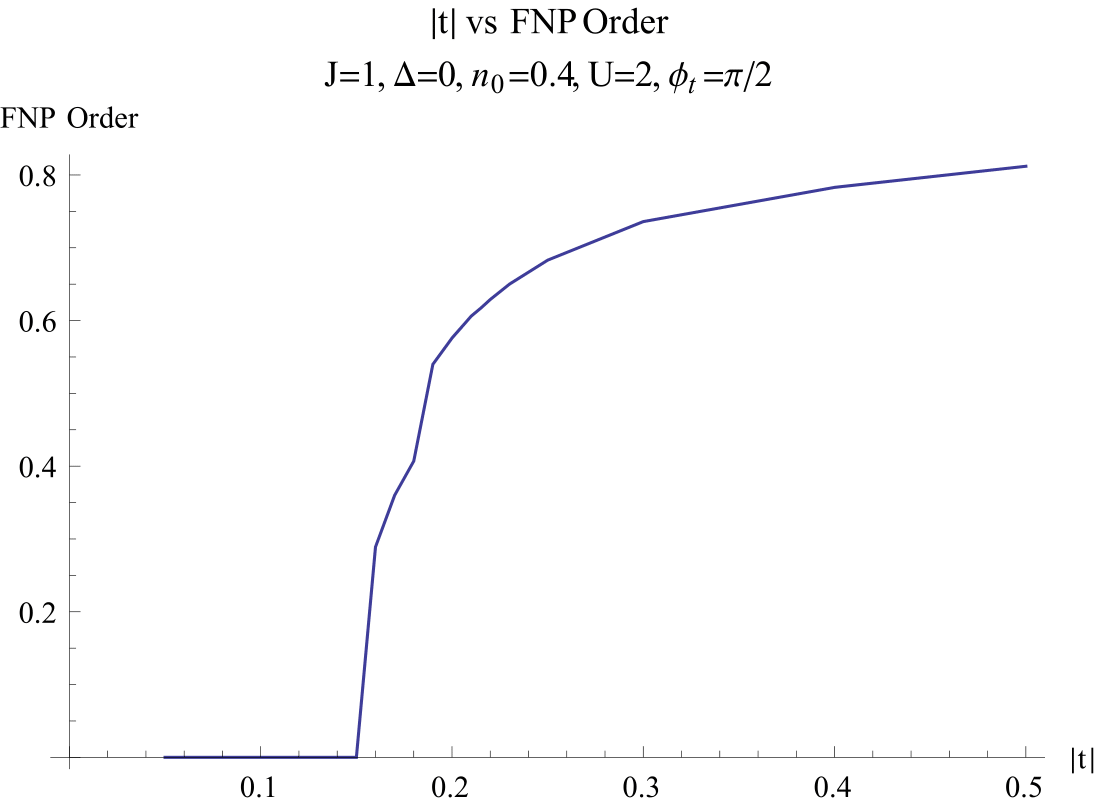}
\caption{\label{fig:phasesC1} For
$U=2$,  $\phi_t=\pi/2$, $J=1$, $\Del=0$, and $n_0=0,0.2,0.4$, we find a
phase transition at $|t|=0.5,0.4,0.15$ respectively.
}
\end{figure}

\begin{figure}[t]
\includegraphics[width=2.4in]{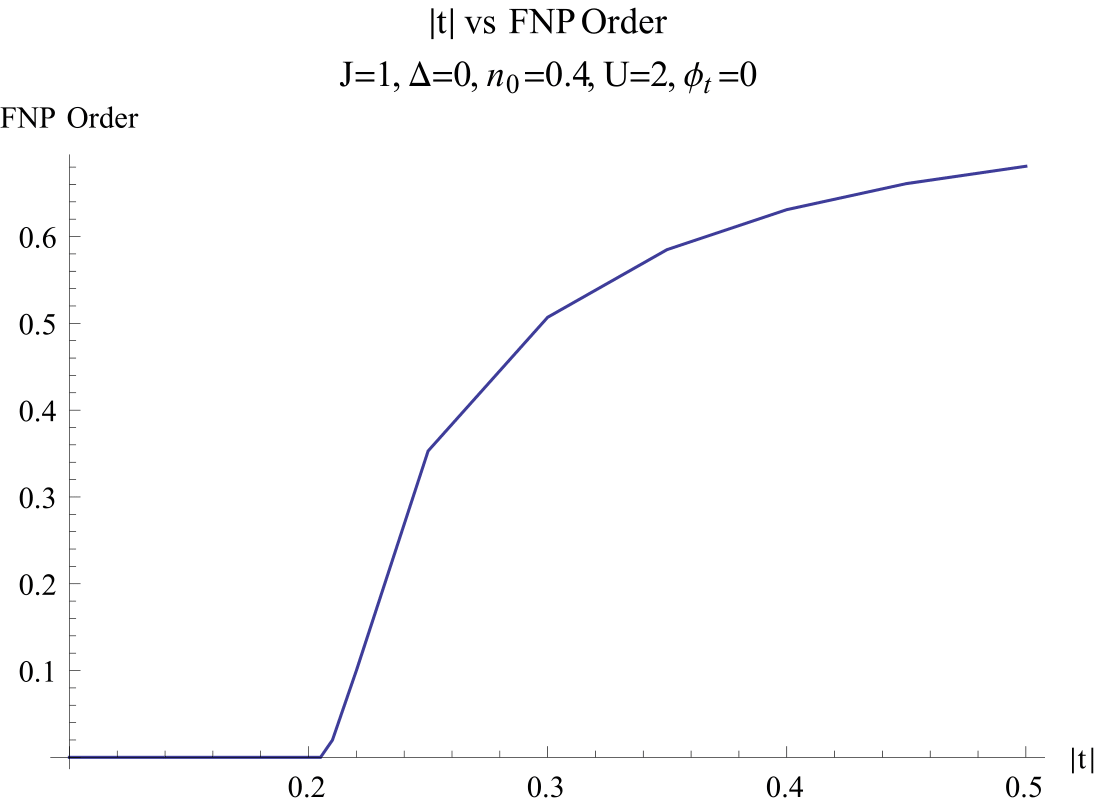}
\caption{\label{fig:phasesC2} For
$U=2$,  $\phi_t=0$, $J=1$, $\Del=0$, and $n_0=0.4$, we find a
phase transition at $|t|=0.2$.
}
\end{figure}

It is straightforward to show that this assumption gives us the following energy expectation value:
\begin{align}
\label{energy function}
\langle H \rangle &= \sum_{i n} 
\Big( U(n-n_0)^2+\Delta \frac{(-1)^n-1}{2}\Big)(r_n^i)^2
\nonumber \\
&+\sum_{i n}J r_{n+2}^i r_n^i \cos(\Delta \theta_n^i+\Delta \theta_{n+1}^i)
\\
+2t\sum_{i,m ,n}& (-1)^n r_{n+1}^i r_{m+1}^{i+1} r_n^i r_m^{i+1} \cos(\Delta \theta_n^i-\Delta \theta_m^{i+1}+\phi_t),
\nonumber 
\end{align}
where $\phi_t$ is the phase of the hopping amplitude
$t=|t|\ee^{\ii \phi_t}$.

We can visualize this as in Fig. \ref{fig:model}, a 2 dimensional classical
system which extends infinitely in one direction ($z$-direction in the case of
Fig. \ref{fig:model}), with interactions between the $\Delta \theta$ sites, and
the strength of those interactions determined by the occupation of the $r$
sites.

We note that the model \eq{energy function} has the FNP
$Z_2^f$ symmetry
\begin{align}
 \Del \th_n^i\to \Del \th_n^i +\pi.
\end{align}
When $J=0$, the model also has the  fermion-number conservation $U(1)$ symmetry
\begin{align}
 \Del \th_n^i\to \Del \th_n^i +\phi.
\end{align}
If we include quantum fluctuations, a $U(1)$ symmetry cannot be spontaneously
broken in 1+1D. So, when $J=0$, the  FNP symmetry $Z_2^f$
cannot be spontaneously broken.  However, when $J\neq 0$, we only have $Z_2^f$
symmetry, which can be  spontaneously broken in 1+1D. Such a $Z_2^f$ symmetry
breaking state is a 1+1D fermionic topologically ordered state, that has a
topological ground state degeneracy on an open line segment.
The $Z_2^f$ symmetry breaking order parameter can be chosen to be
\begin{equation}
\text{FNP order} = \left( \sum_{n=even} r_n^i \right) \left( \sum_{n=odd} r_n^i \right)
\end{equation}

Employing simulated annealing to find the ground state of equation \eq{energy
function} we observe phase transitions in this model.  Choosing $U=2$,
$\phi_t=\pi/2$, $J=1$, $\Del=0$, and $n_0=0,0.2,0.4$ (note that in terms of our energy expression \ref{energy function}  $n_0$ is defined modulo $0.5$), we find a phase
transition at $|t|=0.5,0.4,0.15$ respectively (see Fig. \ref{fig:phasesC1}).
The $Z_2^f$ symmetry breaking appears for large $t$. Also in this case, we observe that the
$Z_2^f$ symmetry breaking ground state is discrete.  

Choosing $U=2$, $\phi_t=0$, $J=1$, $\Del=0$, and $n_0=0.4$ (\ie for real $t$),
we can also find a phase transition at $|t|=0.2$ (see Fig. \ref{fig:phasesC2}).
But in this case, the $Z_2^f$ symmetry breaking ground state is not discrete
and is parametrized by a phase variable $\phi$. More specifically, by observing
the lowest energy configurations of our simulated system we find that the
lowest energy configuration for real $t$ takes the following approximate form:
\begin{align}
\ket{\psi}&=\prod_j \ket{\psi_j} \\
\ket{\psi_j}&=\sum_{n=even} \ii^n E(n)\ket{n} +(-1)^j \ee^{\ii \phi} \sum_{n=odd}  \ii^{n+1} O(n)\ket{n}  
\nonumber 
\end{align}
In this case, after we include the quantum fluctuation of $\phi$, $Z_2^f$
symmetry breaking will be restored.  So the $Z_2^f$ symmetry breaking observed
in the real $t$ case is an artifact of the mean-field theory and there is no $Z_2^f$ symmetry breaking beyond the mean-field theory for $\phi_t=0$.  

\section{Discussion}

The Princeton group has constructed a chain of magnetic iron atoms on
superconducting lead.\cite{NDL1446}.  The iron atoms on the chain are separated
by $\sim$4.2\AA\ and there is also 21\AA-period modulation in the atomic
seperations. The superconducting coherent length of Pb is $\xi=830$\AA, which
is much longer than the total length of the chain which is about  200\AA.  So
the Josephson coupling should have a non-local form $\sum_{i,j>i} J \hat
c_i\hat c_j$, instead of the local form used in \eqn{Hf}. Because of this, the
results in this paper do not apply to Princeton's device.  However, if the iron
chain is much longer than the superconducting coherence length $\xi$ and if the
chain is formed by short segments of length $\xi$ (which can be viewed as
dots), then the chain can be viewed as coupled dots.  In this case, our
approach can be applied to such a system of coupled dots.

\begin{figure}[t]
\includegraphics[width=1.in]{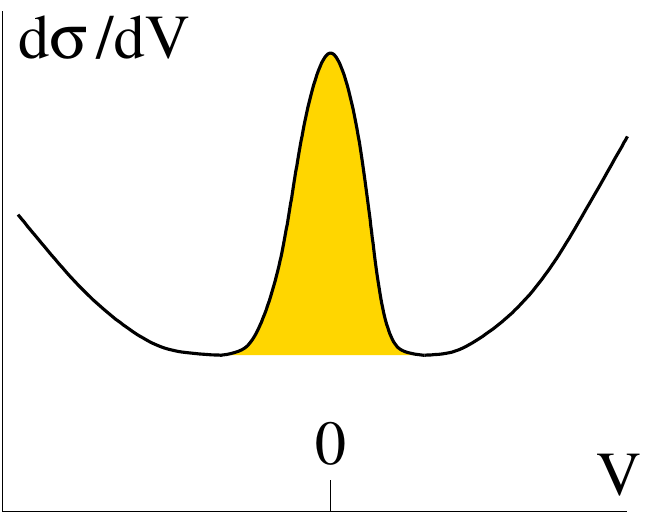}
\caption{\label{tunnel} 
The weight $W$ of the zero-bias tunnelling peak is represented by the shaded area.
}
\end{figure}

The weight $W$ of the zero-bias tunnelling peak into an end of the chain (see
Fig.  \ref{tunnel}) measures the FNP symmetry order parameter.  If we can drive
a zero-temperature phase transition by tuning, for example, the gate voltage
$V$, we expect $W\sim (V-V_c)^\bt$ near the transition with $\bt=1/8$ if there is no other gapless channel on the chain. ($\bt=1/8$ is the
critical exponent of 2D Ising transition).  Such a feature can be used as
a smoking gun to detect the 1+1D fermionic topological order.

JK is supported by NSERC.  XGW is supported by NSF Grant No. DMR-1005541 and
NSFC 11274192.  He is also supported by the BMO Financial Group and the John
Templeton Foundation Grant No. 39901.  Research at Perimeter Institute is
supported by the Government of Canada through Industry Canada and by the
Province of Ontario through the Ministry of Research.

\eject

\bibliography{../../bib/wencross,../../bib/all,../../bib/publst,./local} 

\begin{thebibliography}{29}%
\makeatletter
\providecommand \@ifxundefined [1]{%
 \@ifx{#1\undefined}
}%
\providecommand \@ifnum [1]{%
 \ifnum #1\expandafter \@firstoftwo
 \else \expandafter \@secondoftwo
 \fi
}%
\providecommand \@ifx [1]{%
 \ifx #1\expandafter \@firstoftwo
 \else \expandafter \@secondoftwo
 \fi
}%
\providecommand \natexlab [1]{#1}%
\providecommand \enquote  [1]{``#1''}%
\providecommand \bibnamefont  [1]{#1}%
\providecommand \bibfnamefont [1]{#1}%
\providecommand \citenamefont [1]{#1}%
\providecommand \href@noop [0]{\@secondoftwo}%
\providecommand \href [0]{\begingroup \@sanitize@url \@href}%
\providecommand \@href[1]{\@@startlink{#1}\@@href}%
\providecommand \@@href[1]{\endgroup#1\@@endlink}%
\providecommand \@sanitize@url [0]{\catcode `\\12\catcode `\$12\catcode
  `\&12\catcode `\#12\catcode `\^12\catcode `\_12\catcode `\%12\relax}%
\providecommand \@@startlink[1]{}%
\providecommand \@@endlink[0]{}%
\providecommand \url  [0]{\begingroup\@sanitize@url \@url }%
\providecommand \@url [1]{\endgroup\@href {#1}{\urlprefix }}%
\providecommand \urlprefix  [0]{URL }%
\providecommand \Eprint [0]{\href }%
\providecommand \doibase [0]{http://dx.doi.org/}%
\providecommand \selectlanguage [0]{\@gobble}%
\providecommand \bibinfo  [0]{\@secondoftwo}%
\providecommand \bibfield  [0]{\@secondoftwo}%
\providecommand \translation [1]{[#1]}%
\providecommand \BibitemOpen [0]{}%
\providecommand \bibitemStop [0]{}%
\providecommand \bibitemNoStop [0]{.\EOS\space}%
\providecommand \EOS [0]{\spacefactor3000\relax}%
\providecommand \BibitemShut  [1]{\csname bibitem#1\endcsname}%
\let\auto@bib@innerbib\@empty
\bibitem [{\citenamefont {Mourik}\ \emph {et~al.}(2012)\citenamefont {Mourik},
  \citenamefont {Zuo}, \citenamefont {Frolov}, \citenamefont {Plissard},
  \citenamefont {Bakkers},\ and\ \citenamefont {Kouwenhoven}}]{MZF1236}%
  \BibitemOpen
  \bibfield  {author} {\bibinfo {author} {\bibfnamefont {V.}~\bibnamefont
  {Mourik}}, \bibinfo {author} {\bibfnamefont {K.}~\bibnamefont {Zuo}},
  \bibinfo {author} {\bibfnamefont {S.~M.}\ \bibnamefont {Frolov}}, \bibinfo
  {author} {\bibfnamefont {S.~R.}\ \bibnamefont {Plissard}}, \bibinfo {author}
  {\bibfnamefont {E.~P. A.~M.}\ \bibnamefont {Bakkers}}, \ and\ \bibinfo
  {author} {\bibfnamefont {L.~P.}\ \bibnamefont {Kouwenhoven}},\ }\href
  {\doibase 10.1126/science.1222360} {\bibfield  {journal} {\bibinfo  {journal}
  {Science}\ }\textbf {\bibinfo {volume} {336}},\ \bibinfo {pages} {1003}
  (\bibinfo {year} {2012})}\BibitemShut {NoStop}%
\bibitem [{\citenamefont {Nadj-Perge}\ \emph {et~al.}(2014)\citenamefont
  {Nadj-Perge}, \citenamefont {Drozdov}, \citenamefont {Li}, \citenamefont
  {Chen}, \citenamefont {Jeon}, \citenamefont {Seo}, \citenamefont {MacDonald},
  \citenamefont {Bernevig},\ and\ \citenamefont {Yazdani}}]{NDL1446}%
  \BibitemOpen
  \bibfield  {author} {\bibinfo {author} {\bibfnamefont {S.}~\bibnamefont
  {Nadj-Perge}}, \bibinfo {author} {\bibfnamefont {I.~K.}\ \bibnamefont
  {Drozdov}}, \bibinfo {author} {\bibfnamefont {J.}~\bibnamefont {Li}},
  \bibinfo {author} {\bibfnamefont {H.}~\bibnamefont {Chen}}, \bibinfo {author}
  {\bibfnamefont {S.}~\bibnamefont {Jeon}}, \bibinfo {author} {\bibfnamefont
  {J.}~\bibnamefont {Seo}}, \bibinfo {author} {\bibfnamefont {A.~H.}\
  \bibnamefont {MacDonald}}, \bibinfo {author} {\bibfnamefont {B.~A.}\
  \bibnamefont {Bernevig}}, \ and\ \bibinfo {author} {\bibfnamefont
  {A.}~\bibnamefont {Yazdani}},\ }\href {\doibase 10.1126/science.1259327}
  {\bibfield  {journal} {\bibinfo  {journal} {Science}\ }\textbf {\bibinfo
  {volume} {346}},\ \bibinfo {pages} {602} (\bibinfo {year}
  {2014})}\BibitemShut {NoStop}%
\bibitem [{\citenamefont {Wilczek}(2009)}]{W0914}%
  \BibitemOpen
  \bibfield  {author} {\bibinfo {author} {\bibfnamefont {F.}~\bibnamefont
  {Wilczek}},\ }\href@noop {} {\bibfield  {journal} {\bibinfo  {journal}
  {Nature Phys.}\ }\textbf {\bibinfo {volume} {5}},\ \bibinfo {pages} {614}
  (\bibinfo {year} {2009})}\BibitemShut {NoStop}%
\bibitem [{\citenamefont {Franz}(2010)}]{F1024}%
  \BibitemOpen
  \bibfield  {author} {\bibinfo {author} {\bibfnamefont {M.}~\bibnamefont
  {Franz}},\ }\href@noop {} {\bibfield  {journal} {\bibinfo  {journal}
  {Physics}\ }\textbf {\bibinfo {volume} {3}},\ \bibinfo {pages} {24} (\bibinfo
  {year} {2010})}\BibitemShut {NoStop}%
\bibitem [{\citenamefont {Service}(2011)}]{S1193}%
  \BibitemOpen
  \bibfield  {author} {\bibinfo {author} {\bibfnamefont {R.~F.}\ \bibnamefont
  {Service}},\ }\href@noop {} {\bibfield  {journal} {\bibinfo  {journal}
  {Science}\ }\textbf {\bibinfo {volume} {332}},\ \bibinfo {pages} {193}
  (\bibinfo {year} {2011})}\BibitemShut {NoStop}%
\bibitem [{\citenamefont {{Beenakker}}(2013)}]{B1313}%
  \BibitemOpen
  \bibfield  {author} {\bibinfo {author} {\bibfnamefont {C.~W.~J.}\
  \bibnamefont {{Beenakker}}},\ }\href {\doibase
  10.1146/annurev-conmatphys-030212-184337} {\bibfield  {journal} {\bibinfo
  {journal} {Annu. Rev. Cond. Mat. Phys.}\ }\textbf {\bibinfo {volume} {4}},\
  \bibinfo {pages} {113} (\bibinfo {year} {2013})},\ \Eprint
  {http://arxiv.org/abs/1112.1950} {arXiv:1112.1950} \BibitemShut {NoStop}%
\bibitem [{\citenamefont {Jackiw}\ and\ \citenamefont {Rossi}(1981)}]{JR8181}%
  \BibitemOpen
  \bibfield  {author} {\bibinfo {author} {\bibfnamefont {R.}~\bibnamefont
  {Jackiw}}\ and\ \bibinfo {author} {\bibfnamefont {P.}~\bibnamefont {Rossi}},\
  }\href@noop {} {\bibfield  {journal} {\bibinfo  {journal} {Nucl. Phys. B}\
  }\textbf {\bibinfo {volume} {190}},\ \bibinfo {pages} {681} (\bibinfo {year}
  {1981})}\BibitemShut {NoStop}%
\bibitem [{\citenamefont {Fu}\ and\ \citenamefont {Kane}(2008)}]{FK0807}%
  \BibitemOpen
  \bibfield  {author} {\bibinfo {author} {\bibfnamefont {L.}~\bibnamefont
  {Fu}}\ and\ \bibinfo {author} {\bibfnamefont {C.~L.}\ \bibnamefont {Kane}},\
  }\href@noop {} {\bibfield  {journal} {\bibinfo  {journal} {Phys. Rev. Lett.}\
  }\textbf {\bibinfo {volume} {100}},\ \bibinfo {pages} {096407} (\bibinfo
  {year} {2008})}\BibitemShut {NoStop}%
\bibitem [{\citenamefont {Lee}(2009)}]{L0981}%
  \BibitemOpen
  \bibfield  {author} {\bibinfo {author} {\bibfnamefont {P.~A.}\ \bibnamefont
  {Lee}},\ }\href@noop {} {\  (\bibinfo {year} {2009})},\ \Eprint
  {http://arxiv.org/abs/arXiv:0907.2681} {arXiv:0907.2681} \BibitemShut
  {NoStop}%
\bibitem [{\citenamefont {Lutchyn}\ \emph {et~al.}(2010)\citenamefont
  {Lutchyn}, \citenamefont {Sau},\ and\ \citenamefont {Sarma}}]{LSD1001}%
  \BibitemOpen
  \bibfield  {author} {\bibinfo {author} {\bibfnamefont {R.~M.}\ \bibnamefont
  {Lutchyn}}, \bibinfo {author} {\bibfnamefont {J.~D.}\ \bibnamefont {Sau}}, \
  and\ \bibinfo {author} {\bibfnamefont {S.~D.}\ \bibnamefont {Sarma}},\
  }\href@noop {} {\bibfield  {journal} {\bibinfo  {journal} {Phys. Rev. Lett.}\
  }\textbf {\bibinfo {volume} {105}},\ \bibinfo {pages} {077001} (\bibinfo
  {year} {2010})}\BibitemShut {NoStop}%
\bibitem [{\citenamefont {Choy}\ \emph {et~al.}(2011)\citenamefont {Choy},
  \citenamefont {Edge}, \citenamefont {Akhmerov},\ and\ \citenamefont
  {Beenakker}}]{CEA1119}%
  \BibitemOpen
  \bibfield  {author} {\bibinfo {author} {\bibfnamefont {T.-P.}\ \bibnamefont
  {Choy}}, \bibinfo {author} {\bibfnamefont {J.~M.}\ \bibnamefont {Edge}},
  \bibinfo {author} {\bibfnamefont {A.~R.}\ \bibnamefont {Akhmerov}}, \ and\
  \bibinfo {author} {\bibfnamefont {C.~W.~J.}\ \bibnamefont {Beenakker}},\
  }\href@noop {} {\  (\bibinfo {year} {2011})},\ \Eprint
  {http://arxiv.org/abs/arXiv:1108.0419} {arXiv:1108.0419} \BibitemShut
  {NoStop}%
\bibitem [{\citenamefont {Klinovaja}\ \emph {et~al.}(2013)\citenamefont
  {Klinovaja}, \citenamefont {Stano}, \citenamefont {Yazdani},\ and\
  \citenamefont {Loss}}]{KSY1305}%
  \BibitemOpen
  \bibfield  {author} {\bibinfo {author} {\bibfnamefont {J.}~\bibnamefont
  {Klinovaja}}, \bibinfo {author} {\bibfnamefont {P.}~\bibnamefont {Stano}},
  \bibinfo {author} {\bibfnamefont {A.}~\bibnamefont {Yazdani}}, \ and\
  \bibinfo {author} {\bibfnamefont {D.}~\bibnamefont {Loss}},\ }\href@noop {}
  {\bibfield  {journal} {\bibinfo  {journal} {Phys. Rev. Lett.}\ }\textbf
  {\bibinfo {volume} {111}},\ \bibinfo {pages} {186805} (\bibinfo {year}
  {2013})}\BibitemShut {NoStop}%
\bibitem [{\citenamefont {{Dumitrescu}}\ \emph {et~al.}(2014)\citenamefont
  {{Dumitrescu}}, \citenamefont {{Roberts}}, \citenamefont {{Tewari}},
  \citenamefont {{Sau}},\ and\ \citenamefont {{Das Sarma}}}]{DRT1412}%
  \BibitemOpen
  \bibfield  {author} {\bibinfo {author} {\bibfnamefont {E.}~\bibnamefont
  {{Dumitrescu}}}, \bibinfo {author} {\bibfnamefont {B.}~\bibnamefont
  {{Roberts}}}, \bibinfo {author} {\bibfnamefont {S.}~\bibnamefont {{Tewari}}},
  \bibinfo {author} {\bibfnamefont {J.~D.}\ \bibnamefont {{Sau}}}, \ and\
  \bibinfo {author} {\bibfnamefont {S.}~\bibnamefont {{Das Sarma}}},\
  }\href@noop {} {\  (\bibinfo {year} {2014})},\ \Eprint
  {http://arxiv.org/abs/1410.5412} {arXiv:1410.5412} \BibitemShut {NoStop}%
\bibitem [{\citenamefont {{Peng}}\ \emph {et~al.}(2014)\citenamefont {{Peng}},
  \citenamefont {{Pientka}}, \citenamefont {{Glazman}},\ and\ \citenamefont
  {{von Oppen}}}]{PPG1451}%
  \BibitemOpen
  \bibfield  {author} {\bibinfo {author} {\bibfnamefont {Y.}~\bibnamefont
  {{Peng}}}, \bibinfo {author} {\bibfnamefont {F.}~\bibnamefont {{Pientka}}},
  \bibinfo {author} {\bibfnamefont {L.~I.}\ \bibnamefont {{Glazman}}}, \ and\
  \bibinfo {author} {\bibfnamefont {F.}~\bibnamefont {{von Oppen}}},\
  }\href@noop {} {\  (\bibinfo {year} {2014})},\ \Eprint
  {http://arxiv.org/abs/1412.0151} {arXiv:1412.0151} \BibitemShut {NoStop}%
\bibitem [{\citenamefont {Wen}(1989)}]{W8987}%
  \BibitemOpen
  \bibfield  {author} {\bibinfo {author} {\bibfnamefont {X.-G.}\ \bibnamefont
  {Wen}},\ }\href@noop {} {\bibfield  {journal} {\bibinfo  {journal} {Phys.
  Rev. B}\ }\textbf {\bibinfo {volume} {40}},\ \bibinfo {pages} {7387}
  (\bibinfo {year} {1989})}\BibitemShut {NoStop}%
\bibitem [{\citenamefont {Wen}\ and\ \citenamefont {Niu}(1990)}]{WN9077}%
  \BibitemOpen
  \bibfield  {author} {\bibinfo {author} {\bibfnamefont {X.-G.}\ \bibnamefont
  {Wen}}\ and\ \bibinfo {author} {\bibfnamefont {Q.}~\bibnamefont {Niu}},\
  }\href@noop {} {\bibfield  {journal} {\bibinfo  {journal} {Phys. Rev. B}\
  }\textbf {\bibinfo {volume} {41}},\ \bibinfo {pages} {9377} (\bibinfo {year}
  {1990})}\BibitemShut {NoStop}%
\bibitem [{\citenamefont {Wen}(1990)}]{W9039}%
  \BibitemOpen
  \bibfield  {author} {\bibinfo {author} {\bibfnamefont {X.-G.}\ \bibnamefont
  {Wen}},\ }\href@noop {} {\bibfield  {journal} {\bibinfo  {journal} {Int. J.
  Mod. Phys. B}\ }\textbf {\bibinfo {volume} {4}},\ \bibinfo {pages} {239}
  (\bibinfo {year} {1990})}\BibitemShut {NoStop}%
\bibitem [{\citenamefont {Keski-Vakkuri}\ and\ \citenamefont
  {Wen}(1993)}]{KW9327}%
  \BibitemOpen
  \bibfield  {author} {\bibinfo {author} {\bibfnamefont {E.}~\bibnamefont
  {Keski-Vakkuri}}\ and\ \bibinfo {author} {\bibfnamefont {X.-G.}\ \bibnamefont
  {Wen}},\ }\href@noop {} {\bibfield  {journal} {\bibinfo  {journal} {Int. J.
  Mod. Phys. B}\ }\textbf {\bibinfo {volume} {7}},\ \bibinfo {pages} {4227}
  (\bibinfo {year} {1993})}\BibitemShut {NoStop}%
\bibitem [{Note1()}]{Note1}%
  \BibitemOpen
  \bibinfo {note} {In fact, Majorana fermions had already been found 50 years
  ago in superconductors, but under a different name, the Bogoliubov
  quasiparticles.}\BibitemShut {Stop}%
\bibitem [{\citenamefont {Verstraete}\ \emph {et~al.}(2005)\citenamefont
  {Verstraete}, \citenamefont {Cirac}, \citenamefont {Latorre}, \citenamefont
  {Rico},\ and\ \citenamefont {Wolf}}]{VCL0501}%
  \BibitemOpen
  \bibfield  {author} {\bibinfo {author} {\bibfnamefont {F.}~\bibnamefont
  {Verstraete}}, \bibinfo {author} {\bibfnamefont {J.~I.}\ \bibnamefont
  {Cirac}}, \bibinfo {author} {\bibfnamefont {J.~I.}\ \bibnamefont {Latorre}},
  \bibinfo {author} {\bibfnamefont {E.}~\bibnamefont {Rico}}, \ and\ \bibinfo
  {author} {\bibfnamefont {M.~M.}\ \bibnamefont {Wolf}},\ }\href@noop {}
  {\bibfield  {journal} {\bibinfo  {journal} {Phys. Rev. Lett.}\ }\textbf
  {\bibinfo {volume} {94}},\ \bibinfo {pages} {140601} (\bibinfo {year}
  {2005})}\BibitemShut {NoStop}%
\bibitem [{\citenamefont {Chen}\ \emph {et~al.}(2010)\citenamefont {Chen},
  \citenamefont {Gu},\ and\ \citenamefont {Wen}}]{CGW1038}%
  \BibitemOpen
  \bibfield  {author} {\bibinfo {author} {\bibfnamefont {X.}~\bibnamefont
  {Chen}}, \bibinfo {author} {\bibfnamefont {Z.-C.}\ \bibnamefont {Gu}}, \ and\
  \bibinfo {author} {\bibfnamefont {X.-G.}\ \bibnamefont {Wen}},\ }\href@noop
  {} {\bibfield  {journal} {\bibinfo  {journal} {Phys. Rev. B}\ }\textbf
  {\bibinfo {volume} {82}},\ \bibinfo {pages} {155138} (\bibinfo {year}
  {2010})},\ \Eprint {http://arxiv.org/abs/arXiv:1004.3835} {arXiv:1004.3835}
  \BibitemShut {NoStop}%
\bibitem [{\citenamefont {Chen}\ \emph
  {et~al.}(2011{\natexlab{a}})\citenamefont {Chen}, \citenamefont {Gu},\ and\
  \citenamefont {Wen}}]{CGW1107}%
  \BibitemOpen
  \bibfield  {author} {\bibinfo {author} {\bibfnamefont {X.}~\bibnamefont
  {Chen}}, \bibinfo {author} {\bibfnamefont {Z.-C.}\ \bibnamefont {Gu}}, \ and\
  \bibinfo {author} {\bibfnamefont {X.-G.}\ \bibnamefont {Wen}},\ }\href@noop
  {} {\bibfield  {journal} {\bibinfo  {journal} {Phys. Rev. B}\ }\textbf
  {\bibinfo {volume} {83}},\ \bibinfo {pages} {035107} (\bibinfo {year}
  {2011}{\natexlab{a}})},\ \Eprint {http://arxiv.org/abs/arXiv:1008.3745}
  {arXiv:1008.3745} \BibitemShut {NoStop}%
\bibitem [{\citenamefont {Wang}\ and\ \citenamefont {Wen}(2012)}]{WW1263}%
  \BibitemOpen
  \bibfield  {author} {\bibinfo {author} {\bibfnamefont {J.}~\bibnamefont
  {Wang}}\ and\ \bibinfo {author} {\bibfnamefont {X.-G.}\ \bibnamefont {Wen}},\
  }\href@noop {} {\  (\bibinfo {year} {2012})},\ \Eprint
  {http://arxiv.org/abs/arXiv:1212.4863} {arXiv:1212.4863} \BibitemShut
  {NoStop}%
\bibitem [{\citenamefont {Kitaev}\ and\ \citenamefont {Kong}(2012)}]{KK1251}%
  \BibitemOpen
  \bibfield  {author} {\bibinfo {author} {\bibfnamefont {A.}~\bibnamefont
  {Kitaev}}\ and\ \bibinfo {author} {\bibfnamefont {L.}~\bibnamefont {Kong}},\
  }\href {\doibase 10.1007/s00220-012-1500-5} {\bibfield  {journal} {\bibinfo
  {journal} {Commun. Math. Phys.}\ }\textbf {\bibinfo {volume} {313}},\
  \bibinfo {pages} {351 } (\bibinfo {year} {2012})},\ \Eprint
  {http://arxiv.org/abs/arXiv:1104.5047} {arXiv:1104.5047} \BibitemShut
  {NoStop}%
\bibitem [{\citenamefont {Gu}\ \emph {et~al.}(2010)\citenamefont {Gu},
  \citenamefont {Wang},\ and\ \citenamefont {Wen}}]{GWW1017}%
  \BibitemOpen
  \bibfield  {author} {\bibinfo {author} {\bibfnamefont {Z.-C.}\ \bibnamefont
  {Gu}}, \bibinfo {author} {\bibfnamefont {Z.}~\bibnamefont {Wang}}, \ and\
  \bibinfo {author} {\bibfnamefont {X.-G.}\ \bibnamefont {Wen}},\ }\href@noop
  {} {\  (\bibinfo {year} {2010})},\ \Eprint
  {http://arxiv.org/abs/arXiv:1010.1517} {arXiv:1010.1517} \BibitemShut
  {NoStop}%
\bibitem [{\citenamefont {Kitaev}(2001)}]{K0131}%
  \BibitemOpen
  \bibfield  {author} {\bibinfo {author} {\bibfnamefont {A.~Y.}\ \bibnamefont
  {Kitaev}},\ }\href {\doibase 10.1070/1063-7869/44/10S/S29} {\bibfield
  {journal} {\bibinfo  {journal} {Phys.-Usp.}\ }\textbf {\bibinfo {volume}
  {44}},\ \bibinfo {pages} {131} (\bibinfo {year} {2001})},\ \Eprint
  {http://arxiv.org/abs/arXiv:cond-mat/0010440} {arXiv:cond-mat/0010440}
  \BibitemShut {NoStop}%
\bibitem [{\citenamefont {Sachdev}(2011)}]{Sac11}%
  \BibitemOpen
  \bibfield  {author} {\bibinfo {author} {\bibfnamefont {S.}~\bibnamefont
  {Sachdev}},\ }\href@noop {} {\emph {\bibinfo {title} {Quantum phase
  transitions}}},\ \bibinfo {edition} {2nd}\ ed.\ (\bibinfo  {publisher}
  {Cambridge University Press},\ \bibinfo {address} {Cambridge},\ \bibinfo
  {year} {2011})\BibitemShut {NoStop}%
\bibitem [{\citenamefont {Chen}\ \emph
  {et~al.}(2011{\natexlab{b}})\citenamefont {Chen}, \citenamefont {Gu},\ and\
  \citenamefont {Wen}}]{CGW1128}%
  \BibitemOpen
  \bibfield  {author} {\bibinfo {author} {\bibfnamefont {X.}~\bibnamefont
  {Chen}}, \bibinfo {author} {\bibfnamefont {Z.-C.}\ \bibnamefont {Gu}}, \ and\
  \bibinfo {author} {\bibfnamefont {X.-G.}\ \bibnamefont {Wen}},\ }\href@noop
  {} {\bibfield  {journal} {\bibinfo  {journal} {Phys. Rev. B}\ }\textbf
  {\bibinfo {volume} {84}},\ \bibinfo {pages} {235128} (\bibinfo {year}
  {2011}{\natexlab{b}})},\ \Eprint {http://arxiv.org/abs/arXiv:1103.3323}
  {arXiv:1103.3323} \BibitemShut {NoStop}%
\bibitem [{\citenamefont {Gu}\ and\ \citenamefont {Wen}(2012)}]{GW1248}%
  \BibitemOpen
  \bibfield  {author} {\bibinfo {author} {\bibfnamefont {Z.-C.}\ \bibnamefont
  {Gu}}\ and\ \bibinfo {author} {\bibfnamefont {X.-G.}\ \bibnamefont {Wen}},\
  }\href@noop {} {\  (\bibinfo {year} {2012})},\ \Eprint
  {http://arxiv.org/abs/arXiv:1201.2648} {arXiv:1201.2648} \BibitemShut
  {NoStop}%
\end{thebibliography}%

\end{document}